\documentclass[pre,showpacs,amsmath,amssymb]{revtex4}
\usepackage{bm}
\usepackage{graphics}
\usepackage{epsfig}
\usepackage{graphicx}
\begin{document}

\title{Model of crystal growth with simulated self-attraction}

\author{P. N. Timonin}
\email{timonin@aaanet.ru}

\affiliation{Physics Research Institute at Rostov State University
344090, Rostov - on - Don, Russia}

\date{\today}

\begin{abstract}
The (1+1)-dimensional kinetic model of crystal growth with simulated self-attraction and random sequential or parallel dynamics is introduced and studied via Monte-Carlo simulations. To imitate the attraction of absorbing atoms the probability of deposition is chosen to depend on the number of the nearest-neighbor atoms surrounding the deposited atom so it increases with this number. As well the evaporation probabilities are chosed to roughly account for this self-attraction. The model exhibits the interface depinning transition with KPZ-type roughness behavior in the moving phase. The critical indices of the correlation lengths are $\nu _\parallel   = 0.82 \pm 0.03,{\rm{  }}\nu _ \bot   = 0.55 \pm 0.02$ and the critical index of the growth velocity is $1.08 \pm 0.03$ indicating the new universality class of the depinning transition. The critical properties of the model do not depend on the type of dynamics implemented.
\end{abstract}

\pacs{05.40.-a, 05.70.Fh}

\maketitle

\section{Introduction}
The atomistic kinetic models of crystal growth studied extensively in recent years exhibit the interesting phenomenon of depinning transition between the phases of pinned and moving interface \cite{1}-\cite{5}. The interface roughness also changes drastically at this transition; it is finite in the pinned phase while it diverges in the moving phase of an infinite sample.  The statistics and kinetics of the interface roughness in the moving (rough) phase are universally described by the continuous Kardar-Parisi-Zhang (KPZ) equation \cite{6} (or its limiting case - Edwards-Wilkinson (EW) equation \cite{7}) as it captures the most relevant features of the process forming the interface: surface tension and lateral growth \cite{8}.

Yet KPZ equation can not fully describe the depinning transition as it does not predict the velocity of the interface growth and the roughening transition appears in it only in spatial dimensions $d > 2$ \cite{1}.  So the adequate description of crystal growth warrants the studies of the atomistic kinetic models of Sold-On-Solid (SOS) type devoid of these drawbacks. There are a number of such models imitating surface tension and lateral growth in various ways: via edge of plateau evaporation \cite{3}, polynuclear growth \cite{4} or diffusion \cite{5}. Thus introduced correlations reduce the absolute value of interface roughness and make theoretical results qualitatively similar to the real picture of crystal growth.

At the same time the various SOS models have rather different characteristics of the depinning transition such as critical indices of interface velocity and correlation lengths. Now it is not known how many universality classes of depinning transition can exist. To elucidate this point the studies of other possible SOS models are needed. So here we present one more model which can give a realistic picture of crystal growth. Its motivation relies on the fact that the major role in deposition-evaporation processes belongs to the self-attraction of the atoms of growing crystal. To account for this basic feature the deposition probability of the model should depend on the number of nearest-neighbor atoms surrounding the deposition cell being larger for more populated neighborhood. Apparently, the reverse situation should hold for the evaporation probabilities.
 
Here we consider the $1d$ model with such probabilities of deposition-evaporation processes and random sequential or parallel dynamics. We show that such simulation of self-attraction is sufficient to give a reasonably adequate description of depinning transition. The manifestation of this is, in particular, that the model has the roughness behavior of KPZ type. The critical behavior of the interface velocity and critical indices of the correlation lengths are shown to differ from that of the previously studied models which indicates the new universality class of depinning transition.

\section{Model}
\begin{figure*}
\includegraphics{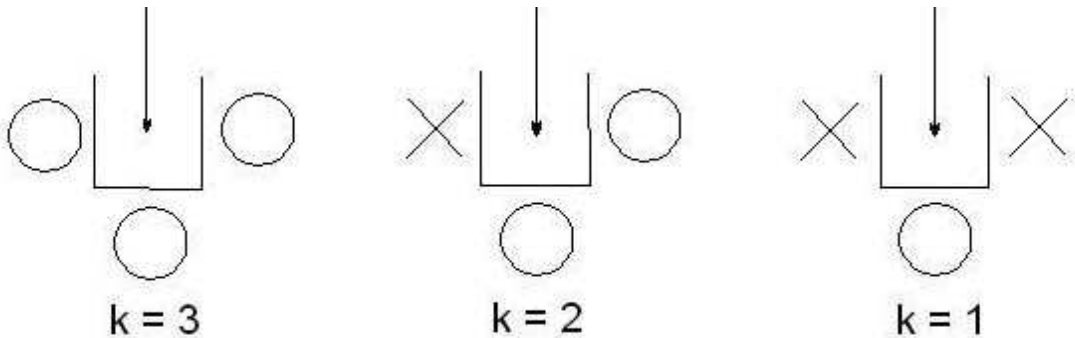}
\caption{\label{Fig.1} Three possible configuration of nearest sites around the deposition cell with three, two and one neighboring atoms. X denotes the empty cell and O shows the atom of crystalline phase.}
\end{figure*}

The crystal growth from the non-crystalline (gaseous or liquid) phase is considered as the result of the deposition and evaporation processes. In the first one the randomly wandering crystal-forming atoms of the non-crystalline phase can occur near the crystal surface and stick to it, while in the second one the atoms evaporate from the crystal surface. The deposition cell in $d = 1$ case can have from one to three nearest neighbors as Fig. \ref{Fig.1} shows. Accordingly to imitate the attraction of the deposited atom to those already belonging to the crystalline phase there should be three different probabilities of absorption $w_k$, ($k = 1, 2, 3$ being the number of nearest neighbors in crystalline phase) such as                                                  
\begin{equation}
1 > w_3 > w_2 > w_1 >0. \label{eq:1}
\end{equation}
                                                                                
Obviously, if $w_1 =1$ then $w_3 = w_2 =1$ and if $w_3 =0$ then $w_1 = w_2 =0$. Here we choose the simplest parametrization, complying with the condition (\ref{eq:1})
 \begin{equation}
w_k = p^{4-k}, \qquad 0 < p < 1 \label{eq:2}
\end{equation}                                                                  
thus reducing the parameter space.
 
To define the evaporation probabilities $u_k$ in configurations of Fig.  \ref{Fig.1} we note that for $k = 3$ the evaporating atom has always three nearest atoms, for $k = 2$ it has two or three neighbors and for $k = 1$ their number can be one, two or three. Then the assumed self-attraction can be roughly simulated by the condition
\begin{equation}
1 > u_1 > u_2 > u_3 >0. \label{eq:3}
\end{equation}
The simple way to comply with it is to assume that no height changes occur with the equal probability $q$ at all $k$. Then we have the evaporation probabilities \begin{equation}
 u_k =1- q - w_k \label{eq:4}
\end{equation}
for the configurations in Fig. \ref{Fig.1}. The positivity of $u_k$ imposes the limitation on the $p$, $q$ values:
\[
                 p + q < 1.	
\]
	Thus the model assumes that the larger number of neighboring atoms the more effectively crystal retains the deposited atom among them and the evaporation takes place more often from the less populated (on the average) neighborhood. Just this one can expect as the result of the interatomic attraction. So the atoms are mainly absorbed and retained at the interface dips while their deposition on the local tops is more rare but the evaporation is more often. On the global scale this acts as the surface tension and induces the lateral growth of interface which result in smoothing its profile which withstand the randomness-induced roughness. Thus we may expect that the model captures the essential features of crystal growth and can give the adequate description of this process.
	
	Apparently there is a vast choice of probabilities complying with Eqs. (\ref{eq:1}, \ref{eq:3}) other than adopted here and some physical arguments can be used to define them. In particular, one may define the evaporation probabilities to depend strictly on the nearest-neighbor number as in the case of the absorption ones. Yet it needs the expansion of the update rules to involve the next-nearest neighbors of the upper atoms of crystal phase. This seems to be not quite necessary for our aim to demonstrate the principal validity of the crystal-growth mechanism based on the self-attraction of the crystal-forming atoms.
	  
	One can choose either random sequential or parallel updates for the model defined by Eqs. (\ref{eq:2}, \ref{eq:4}). Yet random sequential dynamics is more adequate as then at every time step we deal with definite interface profile while in the parallel updates the interference of changes in the nearby sites can not be generally taken into account. Still the alternative parallel updates of the odd and even sites can be implemented in which no interference appears as there are no correlations between the next-nearest sites in the update rules. This type of dynamics should give the results identical to those of random sequential one and our Monte-Carlo simulations corroborate this expectation. Moreover, the critical behavior of the model stays the same under the fully parallel updates; the slight changes appear only in the phase diagram. All the results presented below are obtained with the alternative parallel updates of the odd and even sites.

\section{Monte-Carlo simulations}
\begin{figure*}
\includegraphics{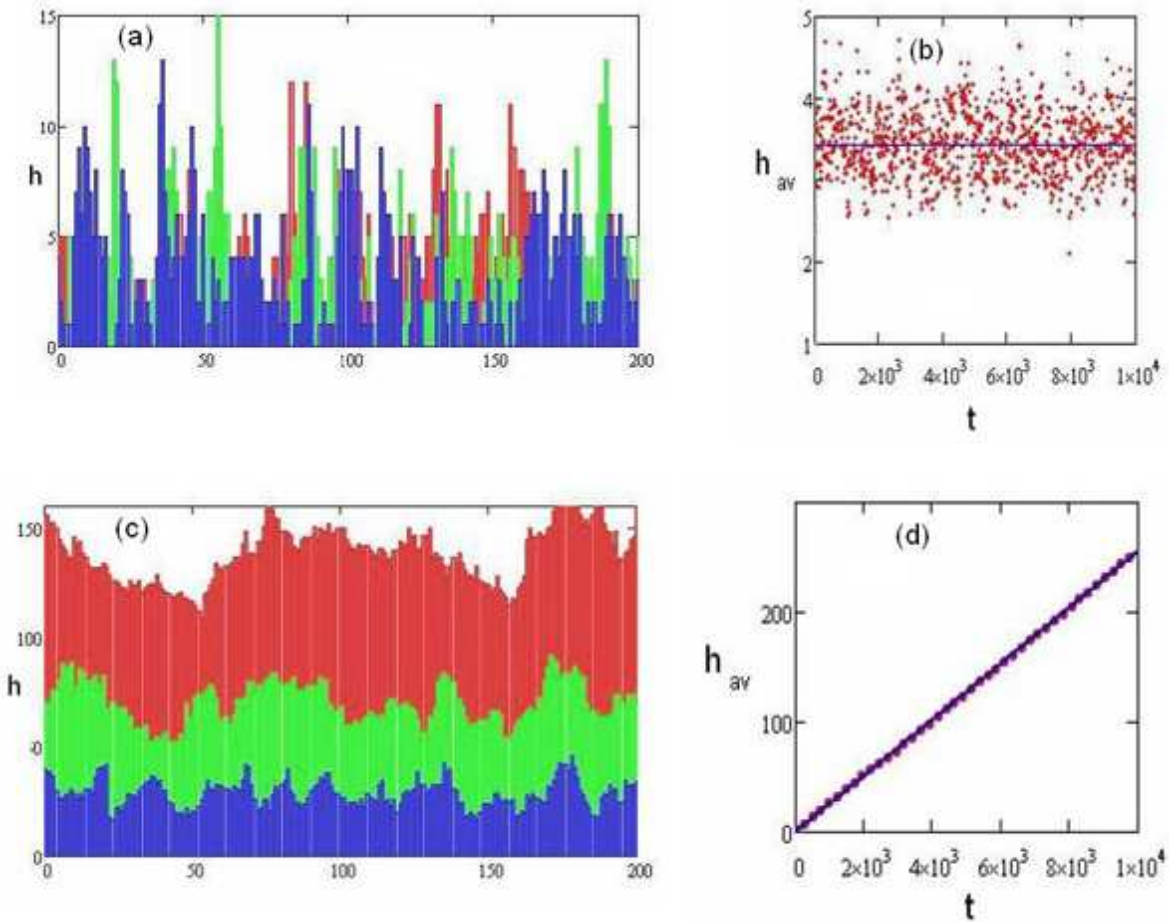}
\caption{\label{Fig.2} The interface profiles and the average heights at $q =0.1$ and $p = 0.67$ (a, b), $p = 0.7$ (c, d) at different times (1000 -blue, 2500 - green, and 5000 - red MC steps). }
\end{figure*}

\begin{figure}
\includegraphics{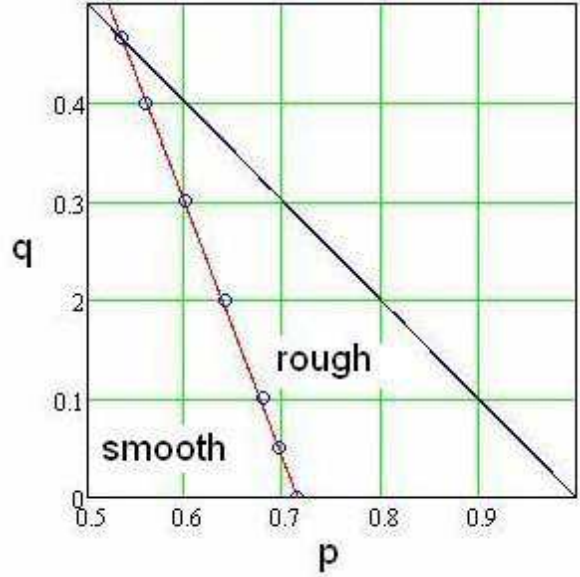}
\caption{\label{Fig.3}.Phase diagram. Circles are the results of MC simulations.}
\end{figure}

\begin{figure}[htp]
\centering
\includegraphics[height=6cm]{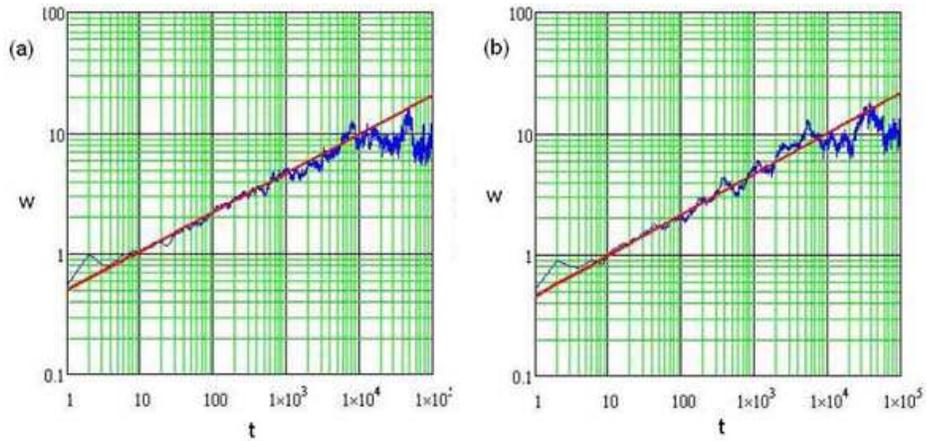}
\caption{\label{Fig.4} The time dependence of roughness for $L = 200$ in the moving phase (a) $p = 0.7$, $q = 0$, (b) $p = 0.7$, $q = 0.1$.  Straight lines correspond to the power law with the exponent 0.333.}
\end{figure}

\begin{figure}[htp]
\centering
\includegraphics[height=6cm]{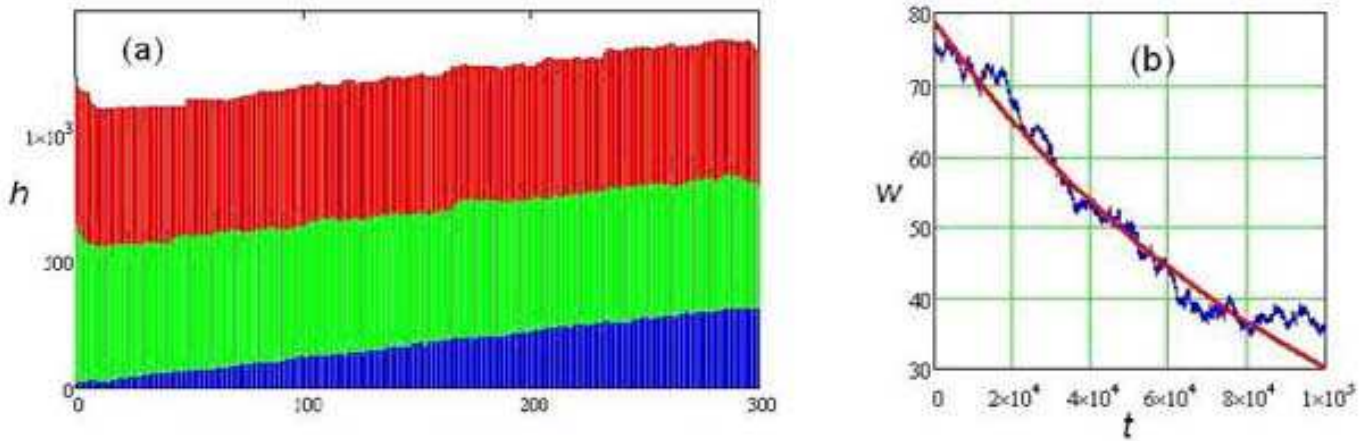}
\caption{\label{Fig.5}(a) Evolution of the interface with initial slope, $q = 0.1$, $p = 0.8$, $L = 300$. The profiles after $t = 100, 3000, 6000$ steps are shown; (b)$ w(t)$, solid line corresponds to Eq.(\ref{eq:5}).}
\end{figure}

\begin{figure}[htp]
\centering
\includegraphics[height=6cm]{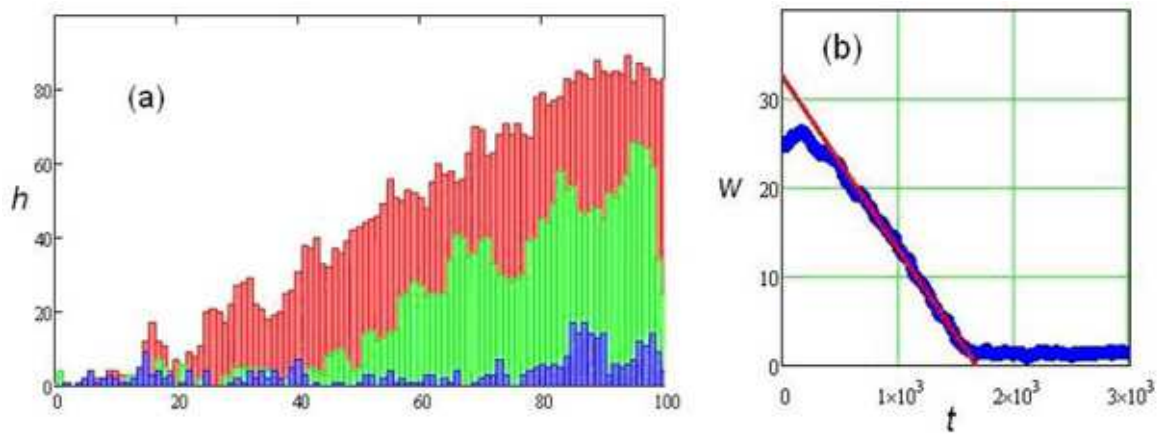}
\caption{\label{Fig.6} Double logarithmic plot $\tau$ vs $p_c - p$. (a) $q = 0.1$, $p = 0.8$. Straight line corresponds to $z = 1.56$, (b) $q = 0$,  $p = 0.8$. Straight line corresponds to $z = 1.96$.}
\end{figure}

The Monte-Carlo simulations were performed on the samples of $L = 100 - 400$ sites with periodic boundary conditions for up to $10^5$ time steps.  The interface profiles emerging at $q =0.1$ and $p = 0.67, 0.7$ and at different times (1000, 2500, and 5000 MC steps) are shown in Fig.\ref{Fig.2}. The appearance of the depinning transition is clearly seen in it. Quantitatively it is manifested in the appearance of the linear in time growth of the average interface height (cf. Fig. \ref{Fig.2}c, \ref{Fig.2}d)
\[
h_{av} \left( t \right) = L^{ - 1} \sum\limits_{n = 1}^L {h_n \left( t \right)} \]
Here $h_n(t)$ is the crystal height of site $n$ at time $t$.
The transition line $p = p_c(q)$ is shown in Fig. \ref{Fig.3}, with good precision it can be approximated by the straight line
\[
 p_c(q) = 0.715 - 0.387q
\]				
The two examples of the time dependence of the interface roughness
\[
w\left( t \right) = L^{-1} \sum\limits_{n = 1}^L {\left| {h_n \left( t \right) - h_{av} \left( t \right)} \right|} 
\]
in the moving (rough) phase are shown in Fig. \ref{Fig.4}. It shows that initial growth of $w(t)$ complies with power low, $w(t) \sim t^{\beta}$. The average growth exponent $\beta$ in it is found to be $\beta = 0.35 \pm 0.04$ for $q = 0, 0.1$ which is close to the KPZ value $\beta_{KPZ} = 1/3$.
To find the dynamical critical index $z$ the simulations with the initial inclined interface $h_n(t = 0) = n$  were performed in the moving phase. In this case initial $w(0) = L/4$ decays approximately exponentially with time to the roughness' saturation value, 
$w\left( \infty  \right) \sim L^\alpha  \ll w\left( 0 \right), \alpha  = \beta z <1$, see Fig. \ref{Fig.5},
\begin{equation}
w\left( t \right) \approx w\left( 0 \right)\exp \left( { - t/\tau } \right) + w\left( \infty  \right) \label{eq:5}
\end{equation}
Here the usual scaling suggests \cite{1}, \cite{2} that $\tau \sim L^z$ sufficiently close to the transition point where spatial correlation length is much greater than $L$. These simulations at $q = 0.1$ and $p = 0.8$ for $L$ = 100, 200, 300, 400 give $z = 1.56 \pm 0.07$ (Fig. \ref{Fig.6}a) which is close to KPZ value $z_{KPZ} = 3/2$ \cite{1}, \cite{2}. For $q =0, p =0.8$ we get $z = 1.96 \pm 0.1$ (Fig. \ref{Fig.6}b) indicating the crossover to the EW behavior with $z =2$ at $p \ge 0.8$. The $\tau$ values in Fig. \ref{Fig.6} are the average over 50-100 trials which appears to be sufficient for the saturation of the standard deviation.

\begin{figure}[htp]
\centering
\includegraphics[height=6cm]{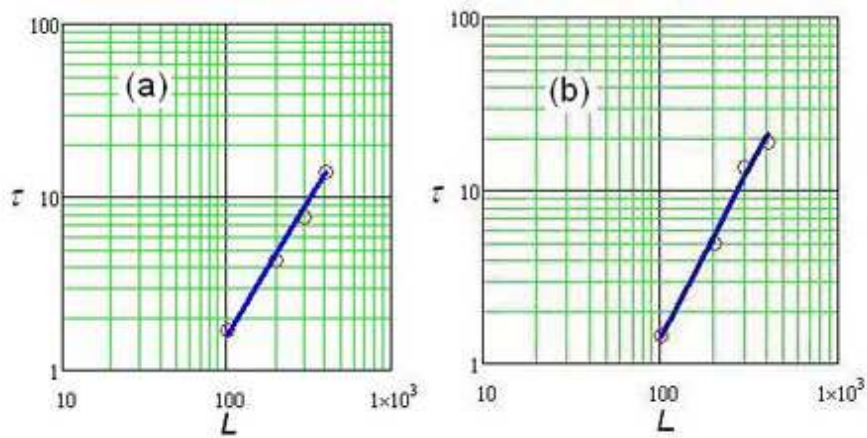}
\caption{\label{Fig.7}(a) Evolution of the interface with initial slope, $q = 0$, $p = 0.68$, $L = 100$. The profiles after $t$ = 150, 750, 1500 steps are shown; (b) $w(t)$, intersection of straight line with the time axis defines the characteristic decay time.}
\end{figure}

We can not reliably confirm that at $q = 0$ in the rough (moving) phase the EW behavior changes to the KPZ one nearer the transition point as it needs the studies of much larger samples. We do not also observe the region of the EW behavior   
$w\left( t \right) \sim t^{1/4} $
at $t$ less than some $t_c$ \cite{9}, cf. Fig. \ref{Fig.4}a. This is probably due to a small $t_c$ value. Yet this crossover can be detected in the critical index of the longitudinal (time) correlation length $\nu_\parallel$ at different $p$ in the pinned phase. To find it we again determine the characteristic time $\tau$ for the vanishing of initial slope but now as a function of $p - p_c$. Fig. \ref{Fig.7} illustrates the results of such simulations.
\begin{figure}[htp]
\centering
\includegraphics[height=6cm]{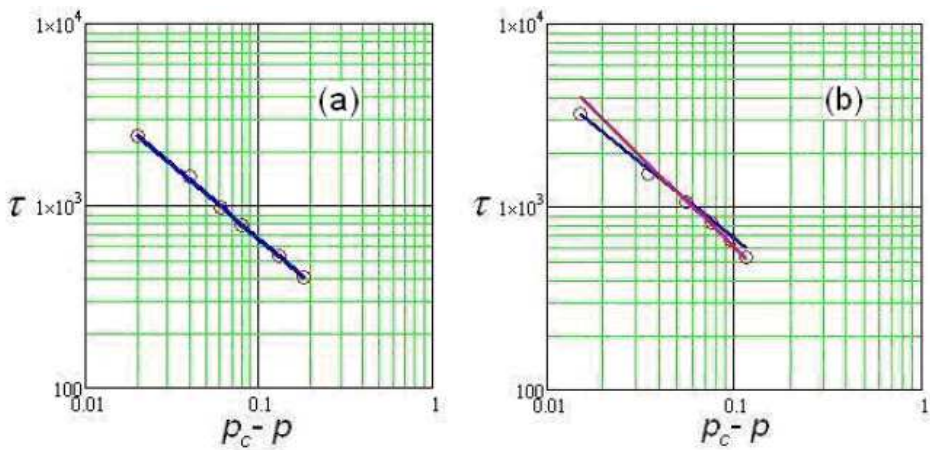}
\caption{\label{Fig.8}(a)Double logarithmic plot $\tau$ vs $L$. (a) $q = 0.1$, $L = 100$. Straight line corresponds to $\nu _\parallel = 0.82 $; (b) $q = 0$, $L = 100$. Straight lines correspond to $\nu _\parallel = 0.83$ and $\nu _\parallel = 1$. $\tau$ is the average over 50-100 trials.}
\end{figure} 
The expected scaling behavior   $\tau  \sim \left( {p_c  - p} \right)^{\nu _\parallel  } $ holds for $q = 0.1$ at all $0.5 < p < p_c = 0.68$ with $\nu _\parallel  = 0.82 \pm 0.03$, cf. Fig. \ref{Fig.8}a. But for $q = 0$ the crossover from the mean-field (EW) $\nu _\parallel = 1$ to the $\nu _\parallel = 0.83 \pm 0.04$ takes place at $p = 0.66$, see Fig. \ref{Fig.8}b. So assuming that in the critical regions of both phases $\nu _\parallel$ and $z$ are the same we may suppose that at $q =0$ the critical behavior of the model also belongs to that of KPZ class. Hence the index of the spatial correlation length is $\nu _ \bot   = \nu _\parallel  /z = 2\nu _\parallel /3 = 0.55 \pm 0.02$. It is just slightly above the mean-field value  $\nu _ \bot   = \nu _\parallel  /z = 1/2$. These $\nu _ \bot$ and $\nu _\parallel$ values are rather low as compared to those known for other (1+1) models. Thus in models of Refs. \cite{3}, \cite{4} they are equal to those of directed percolation $\nu _ {\bot,DP} \approx 1.1$ and $\nu _{\parallel,DP} \approx 1.73$.

\begin{figure}[htp]
\centering
\includegraphics[height=6cm]{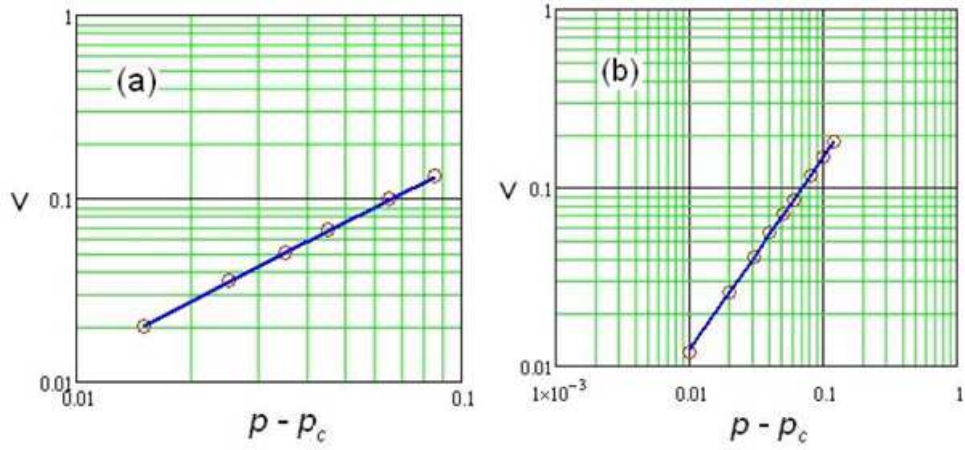}
\caption{\label{Fig.9}Double logarithmic plot $v$ vs $p - p_c$. (a) $q = 0$, $L = 200$. Straight line corresponds to $\gamma = 1.08$; (b)$ q = 0.1$, $L = 200$. Straight line corresponds to $\gamma = 1.08$.}
\end{figure}

 For $q = 0$ and $q = 0.1$ we have also determined in the moving phase the growth velocity index $\gamma$ in the relation $v \sim \left( {p - p_c } \right)^\gamma  $. In both cases it appears to be close to unity, $\gamma = 1.08 \pm 0.03$, see Fig. \ref{Fig.9}. It differs significantly from that of the models in which it can be identified with the index of the longitudinal correlation length of the directed percolation, $\nu _{\parallel ,DP}  \approx 1.73$ \cite{3}, \cite{4}. 

\section{Conclusions}
The present results show that simulation of the attraction between the crystal-forming atoms in the kinetic growth model can provide the realistic picture of crystal growth even in the absence other mechanisms withstanding the interface roughening such as diffusion. The roughening transition in the model belongs to the KPZ class which can be changed to EW one farther from transition point. The depinning transition in it differs in the critical behavior from the previously studied models by the lower indices  $\nu _\parallel$,  $\nu _\bot$  and $\gamma$.

We should also point out that the model lacks the tilting symmetry which KPZ equation has [8], the initial slope in it always vanishes and the interface becomes parallel (on the average) to the initial surface, see Figs. \ref{Fig.5}, \ref{Fig.7}. It seems probable that other SOS models \cite{1}-\cite{5} also have the preferable growth direction. Meanwhile this property becomes lost in the renormalization group procedure for the general macroscopic growth equations with (Gaussian) noise leading invariably to the KPZ equation on the large space and time scales \cite{1}, \cite{8}. This may point out that the Langevin-type stochastic equations can not adequately describe the variety of the depinning transitions and some other more general approaches to their macroscopic characterization should be developed.  

\begin{acknowledgments}
I gratefully acknowledge the useful discussions with V.B. Shirokov and E.D. Gutliansky. 
\end{acknowledgments}

\end{document}